## Journal Name

# ARTICLE

# Low Thermal Conductivity in La-Filled Cobalt Antimonide Skutterudites with Inhomogeneous Filling Factor Prepared Under High-Pressure Conditions

F. Serrano-Sánchez,[a*] J. Prado-Gonjal,[a,h] N. M. Nemes,[b] N. Biskup,[b,c] M. Varela,[b,c] O. J. Dura,[d] J.L. Martínez,[a,e] M.T. Fernández-Díaz,[f] F. Fauth,[g] J.A. Alonso[a]

La-filled skutterudites La$_x$Co$_4$Sb$_{12}$ (x = 0.25, 0.5) have been synthesized and sintered in one step under high-pressure conditions at 3.5 GPa in a piston-cylinder hydrostatic press. The structural properties of the reaction products were characterized by synchrotron x-ray powder diffraction, clearly showing an uneven filling factor of the skutterudite phases, confirmed by transmission electron microscopy. The non-homogeneous distribution of La filling atoms is adequate to produce a significant decrease in lattice thermal conductivity, mainly due to strain field scattering of high-energy phonons. Furthermore, the lanthanum filler primarily acts as an Einstein-like vibrational mode having a strong impact on the phonon scattering. Extra-low thermal conductivity values of 2.39 W m$^{-1}$ K$^{-1}$ and 1.30 W m$^{-1}$ K$^{-1}$ are measured for La$_{0.25}$Co$_4$Sb$_{12}$ and La$_{0.5}$Co$_4$Sb$_{12}$ nominal compositions at 780 K, respectively. Besides that, lanthanum atoms contributed to increase the charge carrier concentration in the samples. In the case of La$_{0.25}$Co$_4$Sb$_{12}$, there is an enhancement of the power factor and an improvement of the thermoelectric properties.

## Introduction

In the modern era, a significant amount of waste heat is lost in energy generation, transport and information processing. Recovery and transformation of that heat into usable electrical energy offers the possibility to improve production efficiencies. Therefore, thermoelectric (TE) materials, which can directly and reversibly convert heat into electrical energy, have been attractive constituents in the research field of sustainable energy. Despite the numerous advantages of thermoelectric generators or cooling devices, such as reliability, stability, endurance or absence of mobile parts, they are not cost-effective so far because of their low conversion efficiency, and are used merely for specific applications.[1–6]

The efficiency in thermoelectric materials is assessed by the figure of merit $ZT = (S^2 \sigma)T/\kappa$, where S stands for the Seebeck coefficient, σ for the electrical conductivity, κ is the thermal conductivity, which contains the sum of the electronic and the lattice contributions and T is the absolute temperature.[3]

Seeking higher thermoelectric performance, many different approaches have emerged during the last decades, such as band engineering, nanostructuration, hierarchical architectures, etc. Among them, the phonon-glass electron-crystal (PGEC) proposed by Slack,[7] consisting in the introduction of "rattling" atoms in structural cages, has had a strong impact on thermoelectric research and is the main motivation driving the investigation of complex materials such as clathrates, Zintl phases and skutterudites.[8]

Nowadays, the research on new thermoelectric materials is focused on lead and telluride-free, environmentally friendly materials that are comprised of earth-abundant elements and suitable for industry and applications.[9] Skutterudites are lead free materials with promising features for thermoelectric applications.[10] These compounds present an intricate CoAs$_3$-like structure defined in the cubic space group *Im-3*, with general formula MPn$_3$, where Pn is a pnictogen (P, As, or Sb) and M a transition metal. There are two main covalent interactions in the structure, between pnictogen atoms, which form rectangular Pn$_4$ rings, and between pnictogen and transition metal atoms forming MPn$_6$ octahedra. Each unit-cell is formed by eight formula units and presents two large voids in 2*a* positions enclosed by MPn$_6$ octahedra, which can be partially filled with guest atoms. Accordingly, as the atomic

[a.] Instituto de Ciencia de Materiales de Madrid, C.S.I.C., Cantoblanco, E-28049 Madrid, Spain.
[b.] Departamento de Física de Materiales, Universidad Complutense de Madrid, E-28040 Madrid, Spain.
[c.] Instituto Pluridisciplinar & Instituto de Magnetismo Aplicado, Universidad Complutense de Madrid, E-28040, Spain.
[d.] Departamento de Física Aplicada, Universidad de Castilla-La Mancha, Ciudad Real, E-13071, Spain.
[e.] ESS Bilbao. Pol. Ugaldeguren III, Pol. A-7B. 48170 Zamudio, Spain.
[f.] Institut Laue Langevin, BP 156X, Grenoble, F-38042, France.
[g.] CELLS–ALBA synchrotron, E-08290 Cerdanyola del Valles, Barcelona, Spain.
[h.] Materials Science Factory, Material Science Institute of Madrid-CSIC, Sor Juana Inés de la Cruz 3, 28049 Madrid, Spain

(*) Electronic mail: fserrano@icmm.csic.es

Electronic Supplementary Information (ESI) available: [SXRD, XRD Rietveld refinements data and thermoelectric properties of different samples batches]. See DOI: 10.1039/x0xx00000x

volume of the elements constituting the crystal structure increases, void size increases.

Most efforts have been dedicated to the study and improvement of the thermoelectric properties of cobalt antimonide skutterudite ($CoSb_3$). This compound displays high carrier mobility, high electrical conductivity and a notably good Seebeck coefficient. Nevertheless, its high thermal conductivity hinders the optimization of the thermoelectric figure of merit. In order to improve its thermoelectric efficiency, three different strategies are commonly employed: (i) modification of the crystal structure by doping (at antimony or cobalt positions),[11–13] (ii) filling in *2a* structural voids[14–17] and (iii) nanostructuration or tuning of the grain morphology through different synthesis methods so that phonon scattering and power factor are enhanced.[10,18–22]

As possible candidates for PGEC, filled $CoSb_3$ have been thoroughly investigated.[23–30] Guest atoms located at *2a* voids are supposed to act as Einstein oscillators, which interact with phonon dispersion processes, thus reducing lattice thermal conductivity.[10,14] Moreover, filler atoms favorably affect the electrical properties of the material as they act as electropositive donors and their effect on the band structure near the Fermi level is minimal.[31,32] Partially filled skutterudites with no substitution in Co or Sb site always give rise to n-type thermoelectrics, as the filler donor effect is not compensated. However, the material can be doped in Co or Sb site for charge compensation. These are p-type materials and the filling limit reaches full occupancy of the *2a* position. Most common filler atoms are alkaline-metals, alkaline-earth metals and rare-earth elements. Jeitschko and Braun carried out the first synthesis and structural study of La filled $Fe_4P_{12}$ and isotypic polyphosphides.[33] Afterwards, many attempts of filling different skutterudites have been carried out. G. S. Nolas *et al.* found that fractional filled lanthanum skutterudites show lower thermal conductivity than fully-filled ones due to disorder effects.[31] In the same vein, the simultaneous inclusion of diverse kinds of elements as fillers, so-called multiple-filled skutterudites, have achieved improved phonon scattering affected by the different resonance frequencies of the rattlers, thus involving a wider range of phonon wavelengths.[18,34–36]

Very recently, a novel approach has been described to efficiently reduce κ, consisting in the uneven filling-fraction distribution in filled skutterudites.[37] The effect of strain-field scattering of high-energy phonons is observed, revealing that an uneven distribution of filling atoms is efficient to further reduce the lattice thermal conductivity of caged crystals: this was realized in filled $La_{0.8}Ti_{0.1}Ga_{0.1}Fe_4Sb_{12}$, synthesized by melt spinning, that showed an ultralow thermal conductivity as a consequence of the segregation of La rich nanodots, randomly distributed in the bulk, thus a wider spectrum of phonons is scattered in this system.[37]

Along these lines, in the present work we describe the direct synthesis of $La_xCo_4Sb_{12}$ (x = 0.25, 0.5) skutterudites under high-pressure (HP) conditions followed by quenching: we discovered that these conditions favor fluctuations in the filling fraction, leading to a "glass-like" ultralow thermal conductivity. By synchrotron x-ray diffraction we have observed the segregation into La-rich and La-poor coexisting skutterudite phases, confirmed by transmission electron microscopy (TEM); the heterogeneous distribution of the rattler element (La) in the sample produces an impressive reduction of the lattice thermal conductivity, mainly due to strain field scattering of high-energy phonons for x = 0.5, along with an enhancement of the power factor for x = 0.25. As a result, a figure of merit ZT=0.51 was obtained for this composition at 657 K.

## Experimental

$La_xCo_4Sb_{12}$ (x= 0, 0.25, 0.5) pellets were prepared by a solid-state reaction under moderate temperature and high pressure conditions. About 1.3 g of a stoichiometric mixture of starting elements La, Co (99 %, ROC/RIC) and Sb (99.5 %, Alfa Aesar) were carefully ground and placed in a niobium capsule (5.5 mm diameter), sealed and introduced inside a cylindrical graphite heater. To prevent oxidation, the capsule was properly manipulated inside an Argon-filled glove-box. Reactions were carried out in a piston-cylinder press (Rockland Research Co.), at a pressure of 3.5 GPa, at 800 ºC for 1 h. Afterwards, the products were quenched to room temperature and the pressure was released. The samples were obtained as hard pellets, which were partially ground to powder for structural characterization, or cut with a diamond saw in a bar shape for transport measurements.

Phase characterization was carried out using X-Ray Diffraction (XRD) on a Bruker-AXS D8 diffractometer (40 kV, 30 mA), run by DIFFRACTPLUS software, in Bragg-Brentano reflection geometry with Cu Kα radiation (λ=1.5418 Å). The data were collected by 0.04 steps over a 2θ range from 10° to 64°. A synchrotron x-ray powder diffraction (SXRPD) study was essential to investigate the structural details of $La_xCo_4Sb_{12}$ (x = 0, 0.25, 0.5) and possible phase segregation, which is not evident with conventional x-ray diffraction techniques. SXRPD patterns were collected in high angular resolution mode (so-called MAD set-up) on the MSPD-diffractometer at ALBA synchrotron in Barcelona, Spain, selecting an incident beam with 29 keV energy (λ= 0.4267 Å). The powdered samples were contained in 0.5 mm diameter quartz capillaries. Temperature dependent SXRPD patterns were collected at 295 (RT), 473, 673, 873 and 1073 K. The structural analysis determined that the samples are mixtures of segregated regions of two phases with different La filling fractions. Throughout the manuscript the samples will, nonetheless, be referred to as their nominal compositions.

The FullProf program[38] was used to study synchrotron data by Rietveld refinement. The line shape of the diffraction peaks was defined by a pseudo-Voigt function. No regions were excluded in the refinement. Complete analysis was carried out refining the following parameters: scale factor, background coefficients, zero-point error, pseudo-Voigt corrected for asymmetry parameters, occupancy of La, Co and Sb, atomic positions and anisotropic displacements for all the atoms.

The analysis of the microstructure and the surface of the pellets at low spatial resolutions was performed by scanning electron microscopy (SEM) in a Hitachi table-top HT-100 microscope. The density of the consolidated pellets was measured using an Archimedes balance ADAM PW184. It was found to be ca. 97% of the crystallographic value for $La_{0.25}Co_4Sb_{12}$ sample and ca. 96% for $La_{0.5}Co_4Sb_{12}$. Transmission electron microscopy (TEM) observations have been carried out in a JEOL 3000F microscope operated at 300 kV, equipped with both Oxford Instruments electron dispersive X-ray (EDX) analyzer ("Inca") and Gatan Enfina electron energy-loss spectroscopy (EELS) analyzer. The later system, used in the scanning TEM (STEM) modes, enables chemical analyses in the nanometric range.

Seebeck coefficient was measured using a commercial MMR-technologies system. Measurements were performed under vacuum ($10^{-3}$ mbar) in the temperature range of 300-800 K. A constantan wire was used as a reference for comparison of bar-shaped samples cut with a diamond saw perpendicular to the pressing direction. Reproducibility was checked with different contacts and constantan wire.

A Linseis LFA 1000 instrument was used to measure the thermal diffusivity ($\alpha$) of the samples over a temperature range of 300 K ≤ $T$ ≤ 800 K by the laser-flash technique. A thin graphite coating was applied to the surface of the pellet to maximize heat absorption and emissivity. The thermal conductivity ($\kappa$) is determined using $\kappa = \alpha\, C_p\, d$, where $C_p$ is the specific heat and $d$ is the sample density. Specific heat was calculated using the Dulong-Petit equation.

## Results and Discussion

### Structural characterization

The reaction products were first characterized by XRD with CuK$_\alpha$ radiation; the patterns for $La_{0.5}Co_4Sb_{12}$ and $La_{0.25}Co_4Sb_{12}$ prepared under high pressure conditions are shown in **Figure 1,** compared with a simulated CoSb$_3$ skutterudite pattern.

A synchrotron x-ray powder diffraction (SXRPD) study was essential to investigate the structural details of $La_xCo_4Sb_{12}$. The crystal structure was defined in the *Im-3* (No 204) space group. Co atoms occupy *8c* Wyckoff positions (1/4, 1/4, 1/4) while Sb atoms occupy *24g* position (0, y, z) and the filler La atoms are located at *2a* sites (0, 0, 0). The high-resolution SXRPD patterns exhibit a splitting of all the diffraction peaks at high scattering angles (see the full patterns and insets in **Figure 2**). The patterns could be modelled as a mixture of two skutterudite phases with very close unit-cell parameters and different filling factors. The occupancy factors of the segregated phases were independently refined; **Table 1** gathers the unit-cell parameters, La occupancy and proportion of each phase in the mixture. The lanthanum content of the La-rich phase is close to x= 0.14 for nominal $La_{0.25}Co_4Sb_{12}$ composition and x= 0.17

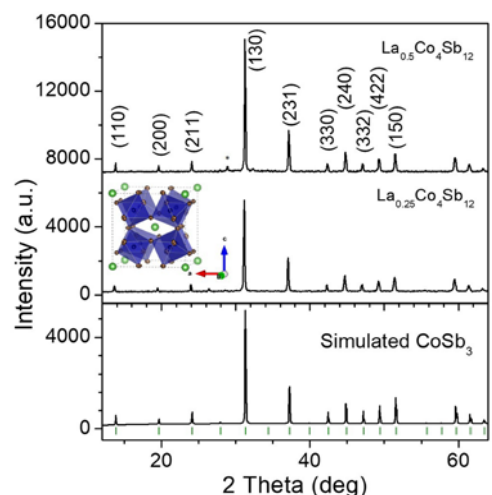

Figure 1. XRD pattern of La$_x$Co$_4$Sb$_{12}$ (x = 0.25 and 0.5) and simulated CoSb$_3$ indexed in a cubic body centered unit cell with a ≈ 9 Å. The star indicates the most intense reflection of Sb impurity. Inset: View of the crystal structure: La (green), Co (blue) and Sb (brown).

for $La_{0.5}Co_4Sb_{12}$. For the La-poor phase, x was refined to 0 and 0.052 for nominal $La_{0.25}Co_4Sb_{12}$ and $La_{0.5}Co_4Sb_{12}$ compositions, respectively (Table 1). The global refined La content is always lower than the nominal stoichiometry, with x=0.11-0.13. The reproducibility of the synthesis was checked for several specimens. Similar average filling levels have been described elsewhere for La-filled skutterudites.[18] A minor impurity of Sb metal was detected and included as a tertiary phase in the refinement of $La_{0.25}Co_4Sb_{12}$; from the scale factors this phase corresponds to a fraction of < 3%. A structure representation of $La_xCo_4Sb_{12}$ is shown as an inset in Fig. 1, highlighting the tilting of the CoSb$_3$ octahedra. The quality of the Rietveld fit from synchrotron powder diffraction data at 295 K is shown in

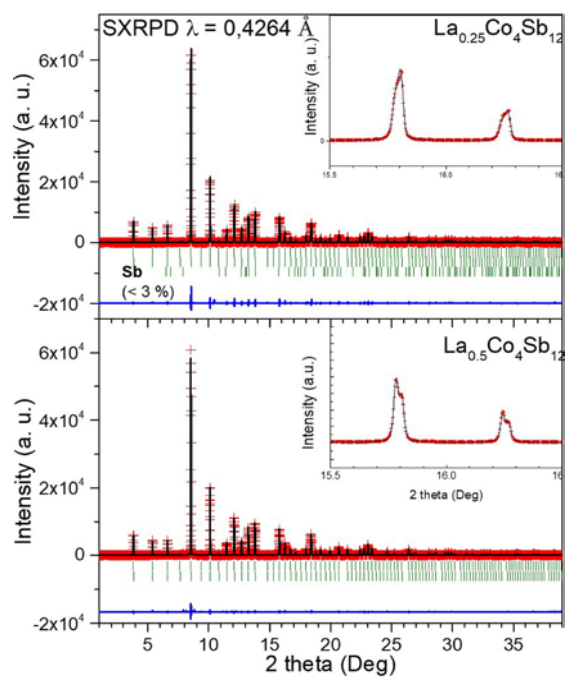

Figure 2. Observed (red crosses), calculated (black full line) and difference (blue line) SXRPD profiles for La$_x$Co$_4$Sb$_{12}$ (x=0.25, 0.5) at 295 K. Insets: close up of the high-angle region showing the splitting of all reflections corresponding to two phases with different filling fraction.

Table 1. Refined structural parameters of La$_x$Co$_4$Sb$_{12}$ (x = 0, 0.25, 0.5) at room temperature. Space group: $Im\bar{3}$

| Nominal Composition | CoSb$_3$ | La$_{0.25}$Co$_4$Sb$_{12}$ | | La$_{0.5}$Co$_4$Sb$_{12}$ | |
|---|---|---|---|---|---|
| Refined Composition | Co$_4$Sb$_{11.49}$ | La$_{0.14}$Co$_4$Sb$_{11.68}$ | Co$_4$Sb$_{11.67}$ | La$_{0.17}$Co$_4$Sb$_{11.64}$ | La$_{0.052}$Co$_4$Sb$_{11.68}$ |
| Phase abundance | 97% (Sb 3%) | 80.5% (Sb 1%) | 18.5% | 63.3% | 36.7% |
| Lattice parameter (Å) | 9.0360(9) | 9.0499(8) | 9.0393(9) | 9.0545(8) | 9.0405(8) |
| Volume (Å$^3$) | 737.80(4) | 741.2(4) | 738.6(2) | 742.3(2) | 738.9(1) |
| $U_{11}$ (Co) / Å$^2$ * | 0.0068(7) | 0.0072(7) | | 0.0063(6) | |
| $U_{12}$ (Co) / Å$^2$ ** | 0.0006(6) | 0.0009(9) | | 0.0003(7) | |
| $y$ (Sb) | 0.3349(1) | 0.3352(3) | | 0.3351(3) | |
| $z$ (Sb) | 0.1578(1) | 0.1581(6) | | 0.1581(4) | |
| Occ. Sb (<1) | 0.966(1) | 0.973(10) | 0.972(4) | 0.971(6) | 0.974(8) |
| $U_{11}$ (Sb) / Å$^2$ *** | 0.0078(2) | 0.0063(6) | | 0.0075(5) | |
| $U_{22}$ (Sb) / Å$^2$ | 0.0108(4) | 0.0115(8) | | 0.0106(7) | |
| $U_{33}$ (Sb) / Å$^2$ | 0.0079(1) | 0.0080(7) | | 0.0077(6) | |
| $U_{23}$ (Sb) / Å$^2$ | 0.0012(4) | 0.0019(5) | | 0.0011(4) | |
| Occ. La (<1) | - | 0.141(13) | - | 0.176(7) | 0.052(13) |
| $U_{11}$ (La) / Å$^2$ **** | - | 0.019(6) | - | 0.008(5) | |
| d Co-Sb (Å) | 2.527(1) | 2.532(5) | 2.528(6) | 2.531(9) | 2.528(1) |
| d$_1$ Sb-Sb (Å) | 2.851(0) | 2.864(7) | 2.860(3) | 2.863(9) | 2.859(5) |
| d$_2$ Sb-Sb (Å) | 2.983(0) | 2.984(4) | 2.979(8) | 2.985(8) | 2.981(2) |
| R$_p$ (%) | 7.53 | 8.67 | | 6.02 | |
| R$_{wp}$ (%) | 11.9 | 10.40 | | 7.67 | |
| R$_{exp}$ (%) | 4.85 | 5.71 | | 5.89 | |
| RBragg (%) | 6.04 | 3.11 | 2.50 | 1.69 | 1.78 |
| $\chi^2$ (%) | 1.21 | 7.26 | | 4.39 | |

Sb at 24$g$, (0,$y$,$z$); Co at 8$c$ (¼,¼,¼); La at 2$a$ (0,0,0)
Anisotropic $U \times 10^4$
Co:* $U_{11} = U_{22} = U_{33}$ ;** $U_{12} = U_{23} = U_{13}$; Sb:*** $U_{12} = U_{13} = 0$; La:****$U_{11} = U_{22} = U_{33}$

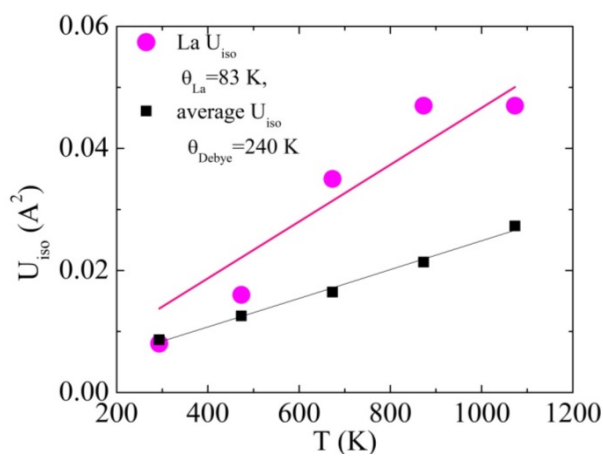

Figure 3. Temperature dependent isotropic / averaged atomic displacement parameters of the majority phase of La$_{0.5}$Co$_4$Sb$_{12}$ sample. The ADP of La, analysed as an Einstein oscillator, gives $\theta_{La}$=83 K. The average ADP, analysed in the Debye model, yields $\theta_{Debye}$=240 K.

Figure 2 for both nominal compositions. Lattice parameters, atomic positions, thermal parameters, bond distances and refinement agreement factors at 295 K are summarized in Table 1. A temperature dependent SXRD study was carried out for La$_{0.5}$Co$_4$Sb$_{12}$. The structural parameters in the 473-1073 K temperature range are listed in Table ESI.1, including the atomic displacement parameters (ADPs), whereas **Figure ESI.1** shows the Rietveld plots above room temperature.

JL Mi *et al*. conducted a thorough analysis of the ADPs based on the Einstein model for different partially filled M$_{0.1}$Co$_4$Sb$_{12}$ (M= La, Ce, Nd, Eu, Yb, Sm) and described a direct relationship between the ADPs of the guest atom and the magnitude of the lattice thermal conductivity.[39]

The Einstein-oscillator energy corresponding to the rattling motion of the La filler ($\theta_{La}$) and possible static displacement (d$_{La}^2$, found to be negligible) were determined by fitting the temperature (T) dependent Atomic Displacement Parameter (ADP or $U_{iso}$ of the La atom) to:

$$U_{iso,La} = \frac{\hbar^2}{2mk_B\theta_{La}} \coth\frac{\theta_{La}}{2T} + d_{La}^2.$$

The Debye-temperature was estimated from the ADPs of the framework atoms (Co and Sb) averaged by site-occupancy (again, static displacement was found to be negligible):

$$U_{iso,<Co/Sb>} = \frac{3\hbar^2 T}{mk_B\theta_D^2}\frac{T}{\theta_D}\left[\int_0^{\frac{\theta_D}{2T}}\frac{x}{e^x-1}dx + \frac{\theta_D}{2T}\right] + d_D^2.$$

Here, this type of analysis of the averaged $U_{iso}$ of the atoms of the host structure, Co and Sb, shows that the Debye temperature is reduced upon La incorporation into the structure, from a value typical for $CoSb_3$ of 262 K to 240 K for $La_{0.5}Co_4Sb_{12}$ and $La_{0.25}Co_4Sb_{12}$, indicative of overall softening of the phonon bands and strong hybridization between the modes of host and "rattler" (**Figure 3**). The disorder parameters are modest in both cases, around 0.03 Å. The temperature dependent $U_{iso}$ of the La rattler can be analyzed as an Einstein oscillator[39,40] and yields an oscillator temperature of around 83 K with no static disorder for $La_{0.5}Co_4Sb_{12}$. An important conclusion of Ref 39 based on a study of various $M_xCo_4Sb_{12}$ samples was that what really reduces the thermal conductivity is simply a large $U_{iso}$ of the "rattler", independent of the Einstein oscillator energies. The room temperature value of $U_{iso}$ of La around 0.008 Å$^2$ is comparable to that found in other studies.[39] We can take this as another indication that, in our samples, κ is driven by the fluctuating filling fraction of the rattler in a nanostructured sample, rather than by changes in the intrinsic phonon modes. Surface texture is evaluated *via* SEM images (**Figure 4**). The nonporous nature of the samples yielded by high-pressure synthesis is assessed in this picture. A compact bulk of apparently single crystalline grains is observed, which are densely packed and sintered to adjacent ones.

The TEM study of the $La_{0.25}Co_4Sb_{12}$ compound is displayed in **Figures 5** and **6**. Figure 5a exhibits a high resolution image of a typical grain. Although the sample is polycrystalline, a dominant orientation close to the [100] zone axis is observed, as shown by the Fourier transform in Figure 5b. Panel 5c shows a few typical crystals of $La_{0.25}Co_4Sb_{12}$, being the grain from Figure 5a highlighted with a red square. The chemical analysis of eight grains visible in Figure 5c was assessed by EDX, revealing that they contain between 1-4% of La, in accordance with the nominal composition. In order to see if this concentration is homogeneously distributed in the grains, we have performed spatially resolved chemical analyses by means of EELS in STEM mode, which allows obtaining compositional information in the 1-10 nm range. EELS data show that the La concentration within the grains is not homogeneous. On the contrary, it varies within the grains, as deduced from the data depicted in Figure 6. Figure 6a shows a low magnification annular dark field (ADF) image of the sample, measured in STEM mode. The region marked with a red rectangle was used to obtain an EEL spectrum image. During the EELS acquisition, the ADF signal was acquired simultaneously to disregard the presence of any major spatial drift (Figure 6b). The Sb $M_{4,5}$, the Co $L_{2,3}$, and the La $M_{4,5}$ edges were measured simultaneously, and variations in the relative Sb / Co / La local intensities were detected. Spectra from two different areas in our EEL spectrum image (depicted with red and green rectangles on panel b)) are shown in corresponding colors in Figure 6c. Variations in the relative edge intensities can be clearly observed. For example, the intense La $M_{4,5}$ edge signal detected in the central area (green rectangle) is almost completely absent in the left (red rectangle) area of the grains. Elemental maps, obtained by integrating the signal under the Sb $M_{4,5}$, Co $L_{2,3}$ and La $M_{4,5}$ edges (after background subtraction using a power law fit) are shown in panels d-f in a false color scale (red = Sb, blue = Co, green = La). While Sb and Co maps are relatively even within the area analyzed (only minor fluctuations are observed), the La map in Figure 6f exhibits a pronounced tendency to segregation, detecting as much as 25 at% of La in the middle of the grain (see the line profile shown in panel 6g obtained along the red line on Figure 6f. A minor decrease of the Co signal is observed in the area where the La species segregate, pointing towards the nucleation of a slightly different phase. However, from these data we cannot fully identify its local composition. The absolute values of the compositional quantification obtained from the EELS analysis have to be taken with a grain of salt due to the significant overlap of the Co $L_{2,3}$ and La $M_{4,5}$ absorption edges used here. In this respect, the EDX analysis is more reliable since it is based on the analysis of high energy Sb $L$, La $L$ and Co $K$ edges. In any case, we can conclude that significant La segregation within lateral length scales of, approximately, 10-30 nm is present. In spite of this local segregation, the average concentration of La in the material remains within the nominal composition range (1.5%).

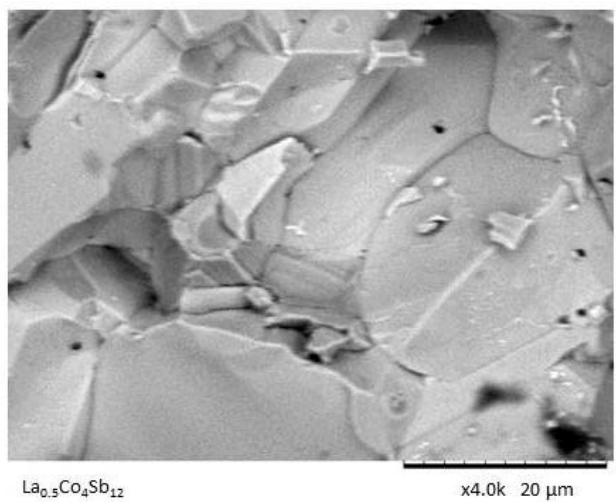

Figure 4. SEM picture of $La_{0.5}Co_4Sb_{12}$ showing a compact, well-sintered morphology.

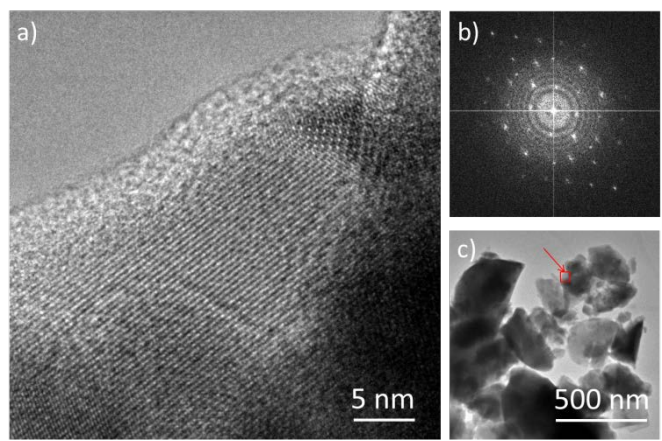

Figure 5. a) High resolution image of a polycrystalline $La_{0.25}Co_4Sb_{12}$ grain. b) Fourier transform of image a) showing a predominant [100] crystal orientation in this grain. c) Overview of grains used for EDX analysis with the grain from panel a) highlighted with a red square.

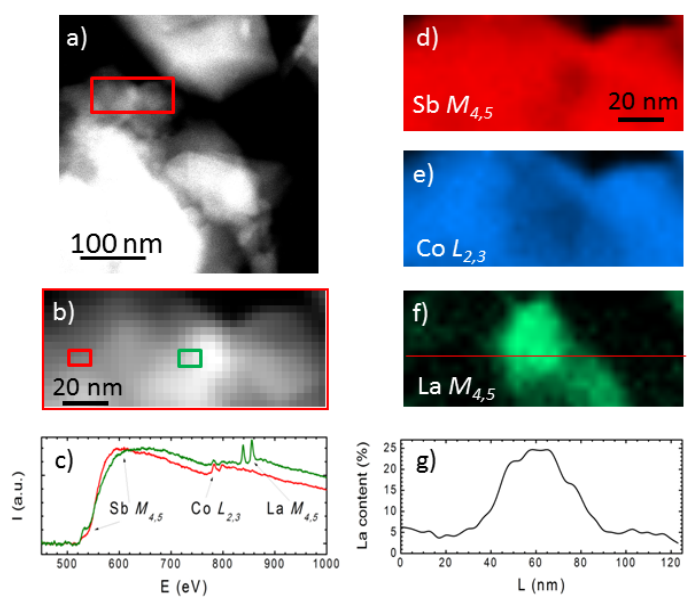

Figure 6. a) Low magnification STEM annular dark field (ADF) image of a few grains, with the region used for EELS analysis highlighted with a red rectangle. b) ADF signal acquired simultaneously with the EELS data. c) Two EEL spectra, taken from the regions depicted with the green and red rectangles marked on panel b). For presentation purposes, the background was subtracted before the Sb $M_{4,5}$ edge using a power law fit. d-f) Elemental maps extracted from the Sb $M_{4,5}$, Co $L_{2,3}$ and La $M_{4,5}$ edges (red, blue and green, respectively). g) Intensity profile of La concentration (atomic %), measured along the direction marked with a red line on the La map in f).

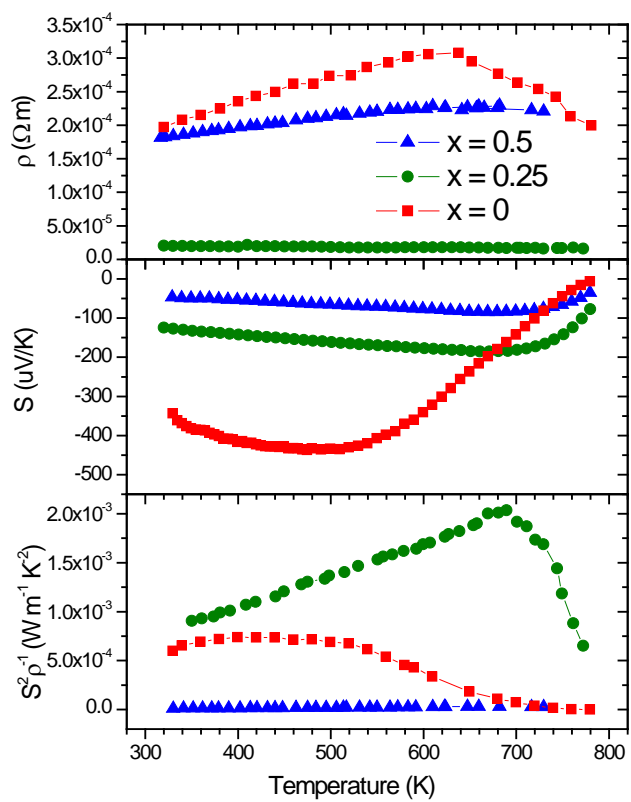

Figure 7. Temperature dependence of the electrical resistivity (ρ), Seebeck coefficient (S) and power factor ($S^2 \rho^{-1}$) of La$_x$Co$_4$Sb$_{12}$ (x = 0, 0.25, 0.5).

**Thermoelectric properties**

The electrical transport properties of the La$_x$Co$_4$Sb$_{12}$ (nominal x = 0.25, 0.50) samples are displayed in **Figure 7** along with those of pure CoSb$_{3-\delta}$ for the sake of comparison, prepared under identical HP conditions.[41] The reproducibility of the transport results was checked in several samples as shown in **Figures ESI.3** and **ESI.4**. La-filled samples display an almost constant value of the electrical resistivity with temperature.

There is partial charge transfer from La to the CoSb$_3$ host, which increases the carrier concentration and thus decreases the resistivity. The resistivity of La$_{0.25}$Co$_4$Sb$_{12}$ slowly decreases with temperature, implying the thermal excitation of carriers. However, La$_{0.5}$Co$_4$Sb$_{12}$ shows metallic behavior: the resistivity slowly increases in this temperature range. This change from semiconductor to metallic behavior was previously described by G.S. Nolas *et al*, who found a similar decrease of resistivity when La fractions of 0.05 and 0.23 were introduced into the structure.[31]

Seebeck coefficient is plotted in the middle panel of Figure 7. Two significant changes take place as a consequence of La filling: The absolute value of the Seebeck coefficient is reduced for the whole temperature range and the maximum is moved towards higher temperatures. The increase in charge carrier concentration, due to lanthanum filling, decreases the Seebeck coefficient, while the effect of minority carriers' excitation at higher temperatures diminishes. The higher the content of La incorporated in the structure, the higher the carrier concentration achieved. Therefore, La$_{0.5}$Co$_4$Sb$_{12}$ shows further reduction and a shift in the maximum absolute Seebeck coefficient. A similar trend was found in the literature for partially lanthanum-filled CoSb$_3$.[31]

The power factor of La$_{0.25}$Co$_4$Sb$_{12}$ is enhanced as a consequence of its low electrical resistivity with a maximum measured value of 2000 μW m$^{-1}$ K$^{-2}$ at 657 K. The reduction of the resistivity is large enough to overcome the fact that the power factor of unfilled CoSb$_3$ is dominated by the Seebeck coefficient, which is diminished in La filled skutterudites.

Temperature dependence of the thermal conductivity is shown in **Figure 8**. Filled and unfilled compounds show similar tendency, as the total thermal conductivity decreases with temperature. This is a typical behavior as lattice thermal conductivity decreases at higher temperatures. Electrical thermal conductivity contribution was calculated using the Wiedemann-Franz law which states $\kappa_e = L\sigma T$, where L is the Lorentz number, σ is the electrical conductivity (σ = ρ$^{-1}$) and T corresponds to the absolute temperature. The value of the temperature-dependent Lorentz number was calculated using the approximation $L = 1.5 + \exp(-|S|/116) * [\![10]\!]^{\wedge}(-8)$.[42]

CoSb$_3$ synthesized by conventional methods exhibits a high thermal conductivity of around 8-10 W m$^{-1}$ K$^{-1}$ at room temperature.[19,43] An exceptionally low thermal conductivity was achieved for a CoSb$_{3-\delta}$ specimen obtained under HP conditions. The total thermal conductivity slightly decreased from 4.3 W m$^{-1}$ K$^{-1}$ at room temperature to 2.6 W m$^{-1}$ K$^{-1}$ at 800 K.[44] Remarkably, the inclusion of La rattling atoms into the skutterudite cavity further reduces the lattice thermal conductivity. Filled La$_{0.25}$Co$_4$Sb$_{12}$ skutterudite displays a reduction in lattice thermal conductivity by approx. 29% at room temperature, whereas for La$_{0.5}$Co$_4$Sb$_{12}$, the reduction is approx. 46% compared to pure CoSb$_{3-\delta}$ at the same temperature (Figure 8a). This drastic change in filled skutterudites is brought about by the rattling of filler atoms in *2a* position, which comes from the strong resonant scattering dependent on the frequency of the rattler. Besides, room-temperature thermal conductivity of 3.0 and 2.3 W m$^{-1}$ K$^{-1}$ for La$_{0.25}$Co$_4$Sb$_{12}$ and La$_{0.5}$Co$_4$Sb$_{12}$ samples, respectively, are really low values even compared to other filled skutterudites.[18,20,37] Nolas *et al.*[31] measured 4 W m$^{-1}$ K$^{-1}$ in La$_{0.23}$Co$_4$Sb$_{12}$ and 7 W m$^{-1}$ K$^{-1}$ lattice thermal conductivity at 400 K in La$_{0.05}$Co$_4$Sb$_{12}$ prepared by a solid state reaction in sealed quartz ampoules at 1073 K and densified by hot pressing at 179 MPa and 873 K. Liu *et al.*[45] reported values between 3.0 and 8 W m$^{-1}$ K$^{-1}$ for a series of La$_x$Co$_4$Sb$_{12}$ (x=0.0-1.0) at 373 K prepared by mechanical alloying and SPS. Park *et al.*[46] found in La$_x$Co$_4$Sb$_{12}$ (x=0.0, 0.1, 0.2, 0.3, 0.4) synthesized by encapsulated melting and hot pressing, thermal conductivity values in the range of 4.0 - 5.0 W m$^{-1}$ K$^{-1}$ at 300 K. The electronic and lattice dependence of the thermal conductivity are represented in the bottom panel of Figure 8a. Lattice thermal conductivity is the main contribution of the total thermal conductivity, which indicates that the process is mainly due to phonon conduction with only an insignificant electronic contribution. The low lattice thermal conductivity values suggest that a filling-fraction fluctuation produced by the phase segregation in the samples has a significant influence on the thermal transport properties. Figure 8b shows the influence of the La filling degree to the total thermal conductivity together with the lattice and electronic contributions at 320 K. It can be observed that the decrease of κ and κ$_L$ is not completely linear.[31,47] Nevertheless, there is a reduction of approx. 17 % in κ and approx. 15 % in κ$_L$ from La$_{0.25}$Co$_4$Sb$_{12}$ to La$_{0.5}$Co$_4$Sb$_{12}$.

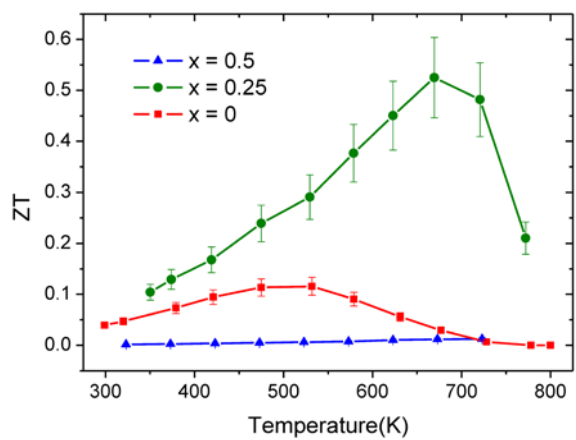

Figure 9. Temperature dependence of the figure of merit, ZT, of La$_x$Co$_4$Sb$_{12}$ (x = 0, 0.25, 0.5). ±15% error bars are included in the graph, based on the accumulation of 5 % error for Seebeck coefficient, 5 % kappa error and 1.5 % for the electrical resistivity error (obtained after 3 repetitions of the measurements).

The increase of the power factor together with the impressive reduction in thermal conductivity in La$_{0.25}$Co$_4$Sb$_{12}$, results in an increase in the figure of merit ZT, compared to HP synthesized CoSb$_{3-\delta}$. In the case of the sample La$_{0.5}$Co$_4$Sb$_{12}$, although the reduction in κ is more substantial, the worsening of the electronic transport properties produces a less-competitive ZT (**Figure 9**). An optimization of the electron counts in La$_x$Co$_4$Sb$_{12}$, while maintaining the reduced lattice thermal conductivity, is expected to produce an enhancement of the TE properties in the future.

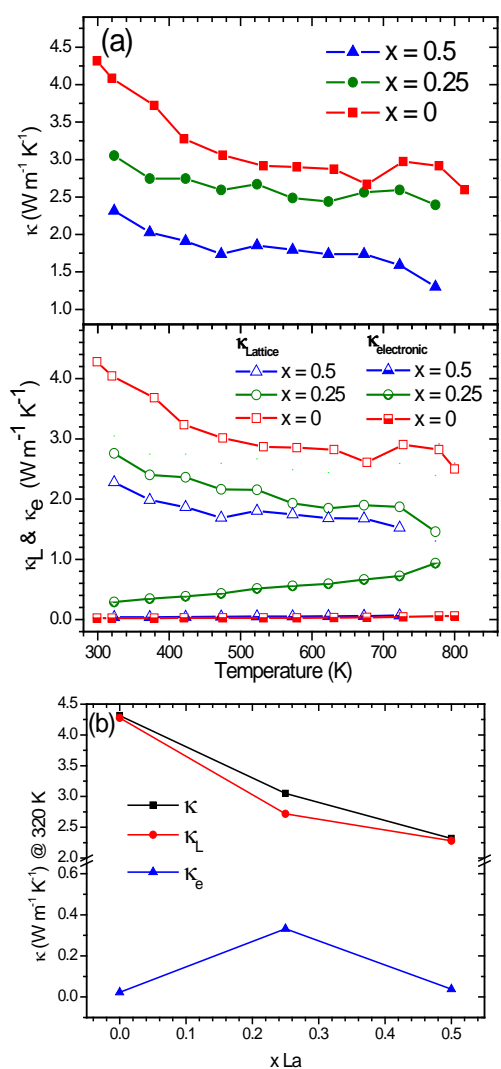

Figure 8. (a) Temperature dependence of the total thermal conductivity (upper panel) and electronic (κ$_e$) and lattice (κ$_L$) contributions (bottom panel) of La$_x$Co$_4$Sb$_{12}$ (x = 0, 0.25, 0.5), (b) total, lattice and electronic thermal conductivity at 320 K vs degree of La filling.

## Conclusion

The synthesis and sintering of La$_x$Co$_4$Sb$_{12}$ (x = 0.25, 0.5) were performed in one step under high pressure conditions, followed by quenching. This procedure drives a segregation of La-filled skutterudite phases, as observed by synchrotron x-ray diffraction experiments and visualized by TEM. The heterogeneous distribution of La rattler elements in the sample produces an impressive reduction of the thermal conductivity, mainly due to strain field scattering of high energy phonons. Additionally, a decrease of the electrical resistivity was observed for La$_{0.25}$Co$_4$Sb$_{12}$, causing the enhancement of its power factor. This result, along with the improvement in thermal conductivity, produces an increase of the figure of merit, ZT, reaching a value of 0.51 at 657 K. These findings may open a new approach to radically decrease the thermal conductivity by promoting the formation of unevenly-filled skutterudites.

## Conflicts of interest

There are no conflicts to declare.

## Acknowledgements

This work was supported by the Spanish Ministry of Economy and Competitivity through grants MAT2013-41099-R, MAT2014-52405-C2-2-R and MAT2015-66888-C3-3-R. JPG would also like to thank this Ministry for granting a Juan de la Cierva fellowship. Financial support from the ERC grant PoC2015-MAGTOOLS is also acknowledged. The authors wish to express their gratitude to ALBA technical staff for making the facilities available for the synchrotron x-ray diffraction experiment.

## Notes and references